\documentclass[12pt]{article}
\usepackage[pctex32]{graphics}
\begin{document}
\vspace{2cm}
\begin{center}
~
\\
~
{\bf  \Large Dual String Description of  Wilson Loop in Non-commutative Gauge Theory}
\vspace{1cm}

                      Wung-Hong Huang\\
                       Department of Physics\\
                       National Cheng Kung University\\
                       Tainan, Taiwan\\

\end{center}
\vspace{2cm}
The Wilson loop in some non-commutative gauge theories is studied by using the dual string description in which the corresponding string is on the curved background with B field.   For the theory in which a constant B field is turned on along the brane worldvolume the Wilson loop always shows a Coulomb phase, as  studied in the previous literature.  We extend the examination to the theory with a non-constant B field, which duals to the gauge theory with non-constant non-commutativity, and re-examine the theory in the presence of a nonzero B field with one leg along the brane worldvolume and other transverse to it, which duals to a non-commutative dipole theory.  The expectation value of the Wilson loop is calculated to the lowest order by evaluating the area of the string worldsheet.  The results show that, while the non-commutativity could modify the Coulomb type potential in IR it may produce a strong repulsive force between the quark and anti-quark if they are close enough.  In particular, we find that there presents a minimum distance between the quarks and that the distance is proportional to the value of the non-commutativity, exhibiting the nature of the non-commutative theory.

\vspace{3cm}
\begin{flushleft}
*E-mail:  whhwung@mail.ncku.edu.tw\\
\end{flushleft}


\newpage
\section{Introduction}
The expectation value of Wilson loop is  one of the most important observations in the gauge theory.   In the AdS/CFT duality [1-3]  it becomes tractable to understand this highly nontrivial quantum field theory effect through
a classical description of the string configuration in the AdS background.  Using this AdS/CFT duality Maldacena [4]  derived for the first time the expectation value of the rectangular Wilson loop operator in the dual string description by evaluating the partition function of a string whose worldsheet is bounded by a loop and along a geodesic on the $AdS_5 \times S^5$ with endpoints on $AdS_5$.    It was found that the interquark potential exhibits the Coulomb type behavior expected from conformal invariance of the gauge theory.

   Maldacena's computational technique has been used to the investigate the Wilson loop in various non-commutative gauge theories. First,  the dual supergravity description of non-commutative gauge theory with a constant non-commutativity was constructed by Hashimoto and Itzhaki [5].   The corresponding string background is the curved spacetime where the constant B field is turned on along the brane worldvolume.  The associated Wilson loop studied by Maldacena and Russo [6] showed that the string cannot be located near the boundary and we need to take a non-static configuration in which the quark-antiquark acquire a velocity on the non-commutative space.  The result shows that the interquark potential exhibits the Coulomb type behavior as that in commutative space.  In [7] Alishahiha, Oz, and Sheikh-Jabbari had found the most general dual supergravity of Dp branes in the presence of the constant B field. They calculated the interquark potential therein and showed that it has the same behavior as that in the $B=0$, up to a constant which is proportional to $B$. 

In section II we will use AdS/CFT correspondence to investigate the Wilson loop on the non-commutative gauge theory with a non-constant non-commutativity.  The dual supergravity is the Melvin-Twist deformed $AdS_5\times S^5$ background which was first constructed by Hashimoto and Thomas [8].\footnote {The extensions to the more general backgrounds, such as those on the deformed Dp brane background, were constructed in [9].} The corresponding string is on the curved background where the non-constant B field is turned on along the brane worldvolume.  After analyzing the Nambu-Goto action of the classical string configuration we see that, while the non-commutativity could modify  the Coulomb type potential in IR it may produce a strong repulsive force between the quark and anti-quark if they are close enough.  In particular, we show that there presents a minimum distance between the quarks, which is proportional to the value of the non-commutativity.

  Another interesting non-commutative gauge theory is the non-commutative dipole field theory [10-12].  The dual supergravity background had been found in [13].  In [14] Alishahiha and Yavartanoo found the most general dual supergravity of Dp branes in the presence of a nonzero B field with one leg along the brane worldvolume and other transverse to it, which duals to a non-commutative dipole theory.  They had investigated the associated Wilson loop and found that when the distance between quark and anti-quark is much bigger then their dipole size the energy will show a Coulomb type behavior with a small correction form the non-commutativity. 

 In section III the expectation value of the Wilson loop in a non-commutative dipole gauge theory, which had been derived in [14], will be re-analyzed. The new property we found is that there is  a strong repulsive force between the quark and anti-quark if they are close enough and there exists a minimum distance between the quarks.  

 Note that the non-commutative deformation of the Maldacena-Nunez supergravity solution [15] was constructed by Mateos et.al. [16].  By calculating the interquark potential via the Wilson loop they found that the corresponding gauge theory  shows confinement in the IR and strong repulsion at closer distances.  However, there does not exist a  minimum distance between the quarks.  The existence of a minimum distance between the quark and antiquark had been found by us in a previous paper [17], in which the dual string are propagating on the magnetic Melvin field deformed spacetime.


\section{Wilson Loop in Non-commutative Gauge Theory }
The geometry of the Melvin-twist deformed $AdS_5\times S^5$ background we considered are described by [8]
$$ds_{10}^2 = U^2\left(- dt^2+ dr^2+{r^2 d\phi^2 + dz^2\over 1+ B^2r^2U^4}\right)+ {1\over U^2} \left(dU^2+ U^2d\Omega_5^2\right). \eqno{(2.1)}$$
$$e^{2\Phi}= {1 \over  1+ B^2r^2U^4},~~~~~C_{tr}= B^2rU^4, ~~~~C_{tr\phi z}= {rU^4 \over  1+ B^2r^2U^4}$$
$$B_{\phi z}= {Br^2U^4 \over  1+ B^2r^2U^4}.\hspace{7.5cm} \eqno{(2.2)}$$
\\
From (2.1) we see that, as the spacetime  $AdS_5$ has been deformed the original isometry group of SO(2.4), which corresponding the conformal symmetry, is broken.  As discussed in [8] the supersymmetry is broken in the supergravity dual.  The D3 worldvolume described by the coordinate in $r$, $\phi$, $z$ becomes non-commutative as $B_{\phi z}\ne 0$.  Also, as the NS-NS field $B_{\phi z}$ depends on a worldvolume coordinate $r$ the dual non-commutative gauge theory has a space-dependent non-commutativity [8].

Note that the Melvin background described in (2.1) and (2.2) was constructed through the T duality and making a twist.   In the original literature [18], after a twist the Kaluza-Klein reduction is used to find the lower dimensional theory in which there will appear an EM Melvin filed.  However, the Melvin Universe found in [8] does not use the Kaluza-Klein reduction and there have not EM Melvin filed.  The Melvin Universe is thus supported by the flux of the NSNS B-field.

Following the Maldacena's computational technique the Wilson loop of a quark anti-quark pair is calculated from a dual string.  The string lies along a geodesic with endpoints on the deformed $AdS_5$ boundary representing the quark and anti-quark positions.  The ansatz for the background string we will use is 
$$ t=\tau,~~~z=\sigma,~~~\phi = v\tau,~~~U=U(\sigma),\eqno{(2.3)}$$
and rest of the string position is constant in $\sigma$ and $\tau$.

   As that discussed by Maldacena and Russo [6],  the string cannot be located near the boundary and we need to take a non-static configuration in which the quark-antiquark acquire a velocity on the non-commutative space.  In the case with a constant non-commutativity the quark-antiquark will move along a direction with constant coordinate $x=x_0$ [8].  However, in our case as the NS-NS field $B_{\phi z}$ described in (2.2) depends on a worldvolume coordinate $r$ the quark-antiquark will rotate along a constant coordinate  $r=r_0$ with a velocity $v$.  The physical interpretation is somewhat analogous to the situation when a charged particle enters a region with a radius-dependent magnetic field.   In this case there will have a relation between the three parameter: the radius of the circular orbit $r_0$, the velocity $v$, and strength of the field $B$.  The relation will be investigated in this section.  

 After choosing $r=r_0$ the Nambu-Goto action becomes
$$S= {1\over 2\pi}\int d\sigma d\tau \left(\sqrt{- det g} +B_{\mu\nu}\partial_\tau X^{\mu}\partial_\sigma X^{\nu}\right)~~\hspace{7cm}$$
$$={T\over 2\pi}\int d\sigma \left(\sqrt{\left((\partial_\sigma U)^2 +{U^4\over 1+ B^2r_0^2U^4}\right)\left(1- {r_0^2v^2\over 1+ B^2r_0^2U^4} \right)}  ~+{v  Br_0^2U^4\over 1+ B^2r_0^2U^4} \right),\eqno{(2.4)}$$
in which $T$ denotes the time interval we are considering and we have set $\alpha'=1$.  As the associated Lagrangian $({\cal L})$ does not explicitly depend on $\sigma$ the relation $(\partial_\sigma U){\partial{\cal L}\over \partial(\partial_\sigma U)} - {\cal L}$ will be proportional to an integration constant.  This implies the following relation
$$ {U^4(1-r_0^2v^2)\over \sqrt{((\partial_\sigma U)^2 +U^4)(1-r_0^2v^2)}} - v B r_0^2 U^4  = {U_0^4(1-r_0^2v^2)\over \sqrt{U_0^4(1-r_0^2v^2)}} - v Br_0^2 U_0^4,\eqno{(2.5)}$$
in which we will put the quark at place $z=\sigma =-L/2$ and the anti-quark at $z=\sigma = L/2$.  Thus at $z=\sigma =0$ we have the relations $U\equiv U_0$ and $\partial_\sigma U = 0$.  Note that the above relation is the leading approximation of small non-commutativity, i.e. $B \ll 1$. 

 From the above relation we can first find the function form of $\partial_\sigma U$
$$\partial_\sigma U={U^2~ \sqrt{(U^4-U_0^4)\left(\sqrt{1-r_0^2v^2}-2vB r_0^2\right)}\over\sqrt{U_0^4\sqrt{1-r_0^2v^2}+2vB r_0^2 U_0^2(U^4-U_0^4)}}.\eqno{(2.6)}$$
Using the above relation we have a result   
$${L/2} = \int_0^{L/2} d\sigma = \int_{U_0}^\infty dU (\partial_\sigma U)^{-1}\hspace{6.6cm}$$
$$={1\over 4U_0}{(1-r_0^2v^2)^{1/4}\over \sqrt{(1-r_0^2v^2)^{1/2}+2vB r_0^2}}\int_0^1 {d x\over x^{3/4}}{\sqrt{x+ {2vB r_0^2U_0^2\over \sqrt{1-r_0^2v^2}}(1-x)}\over \sqrt{1-x}}.\eqno{(2.7)}$$
For a clear illustration we show in figure 1 the function $L(U_0)$ which is that by performing the numerical evaluation of (2.7) for the cases of $B=0$, $B=0.05$ and $B=0.3$, respectively, and $r_0=1$, $v=0.8$.  
\\
\\
\scalebox{1}{\hspace{5cm}\includegraphics{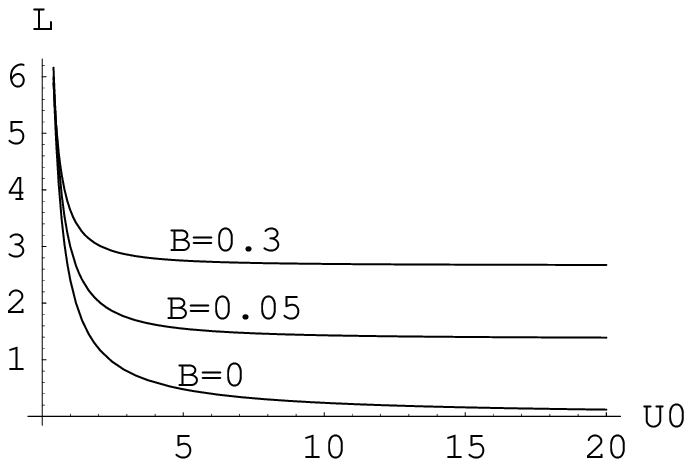}}
\\
{\hspace{3cm} {\it Figure 1.  The function $L(U_0)$ for the cases of $B=0$, $B=0.05$ and $B=0.3$ respectively.   We see that there presents a minimum distance between the quarks and that the minimum distance  is proportional to the strength of the non-commutativity $\sqrt B$. }
\\
\\
Figure 1 shows that there presents a minimum distance between the quarks and that the minimum distance  is proportional to the strength of the space non-commutativity $\sqrt B$.   This property could be read from the following analyses.

  Using (2.7) we can find 
$$L \approx  {1\over 2U_0}{(1-r_0^2v^2)^{1/4}\over \sqrt{(1-r_0^2v^2)^{1/2}+2vB r_0^2}}\left[4 ~ {\sqrt{2vB r_0^2U_0^2}\over (1-r_0^2v^2)^{1/4}}+{1\over 2}log\left({2vB r_0^2U_0^2\over \sqrt{1-r_0^2v^2}}\right) {(1-r_0^2v^2)^{1/4}\over\sqrt{2vB r_0^2U_0^2}}+ \cdot\cdot\cdot\right] $$
$$= {\sqrt{8vr_0^2} \over (1-r_0^2v^2)^{1/4}}~\sqrt B ~ +
{1\over 4U_0}{1\over\sqrt{2vB r_0^2U_0^2}}
log\left({2vB r_0^2U_0^2\over \sqrt{1-r_0^2v^2}}\right) {(1-r_0^2v^2)^{1/2}\over \sqrt{(1-r_0^2v^2)^{1/2}+2vB r_0^2}}+ \cdot\cdot\cdot,\eqno{(2.8)}$$
which is the approximation of the large $U_0$, i.e. at short distance. From above result we have found a minimum distance 
$$L_0\equiv {\sqrt{8vr_0^2} \over (1-r_0^2v^2)^{1/4}}~\sqrt B,\eqno{(2.9)}$$
which is proportional to the strength of non-commutativity $\sqrt B$. This means that the non-constant non-commutativity in there could produce a strong repulsive force between the quark and anti-quark.  Note that in a previous paper [16] Mateos et.al. had found that the non-commutative deformation has an effect to produce a strong repulsion at closer distances.  However, there does not exist a minimum distance between the quarks. The quark-antiquark potential in the dual non-commutative deformation of the Maldacena-Nunez supergravity solution shows confinement in the IR [16].  However, the confinement does not show in the dual background (2.1).  Let us analyze the behavior in below.

First, form (2.7) we can find 
$$L \approx {1\over 2U_0}{\Gamma(3/4)\over\Gamma(5/4) }{(1-r_0^2v^2)^{1/4}\over \sqrt{(1-r_0^2v^2)^{1/2}+2vB r_0^2}}+ \cdot\cdot\cdot,\eqno{(2.10)}$$
which is the approximation of the small $U_0$, i.e. at large distance.  Next, using (2.6) we can evaluate the interquark potential $H$ form the Nambu-Goto action (2.4).  The formula is
$$H = {1\over\pi}{(1-r_0^2v^2)^{3/4}\over \sqrt{(1-r_0^2v^2)^{1/2}+2vB r_0^2}}\int_{U_0}^{\infty} dU {U^2\over \sqrt{U^4-U_0^4}}\hspace{7cm}$$
$$ + {1\over\pi}{vBr_0^2\over \sqrt{(1-r_0^2v^2)^{1/2}+2vB r_0^2}}\int_{U_0}^{\infty} dU{U^2 \sqrt{U_0^2 (1-r_0^2v^2)^{1/2}+ 2vBr_0^2(U^4-U_0^4)U_0^2}\over\sqrt{U^4-U_0^4}} $$
$$\approx {U_0\over \pi}{(1-r_0^2v^2)^{3/4}\over \sqrt{(1-r_0^2v^2)^{1/2}+2vB r_0^2}}\left[\int_1^\infty dy \left({y^2\over \sqrt{y^4-1}}-1\right)y^\epsilon-1\right],\hspace{4cm}\eqno{(2.11)}$$
which is the leading approximation of small non-commutativity, i.e. $B \ll 1$.  Note that as the original integration is a divergent quantity we have followed the prescription of Maldacena [4] to multiply the integration by $y^\epsilon$ and subtraction the regularized mass of W-boson to find the finite result.  After the calculation the potential $H$ becomes
$$H \approx  -{\sqrt{1-v^2r_0^2}+ 2vBr_0^2\over2\pi}{\Gamma(3/4)\over\Gamma(5/4)}{1\over L}\equiv H_0{1\over L}, \eqno{(2.12)}$$
in which we have used the relation (2.10).  Thus the non-commutativity will  decrease the Coulomb type potential in IR.   (Note that due to the log term in (2.8) we could not express $U_0$ as an analytic function of $L$.  Thus it is difficult to find a simple expression of the interquark potential when $L \rightarrow L_0$.)

As has mentioned that there will have a relation between the three parameter: the radius of the circular orbit $r_0$, the velocity $v$, and strength of the field $B$. To determine the relation let us regard the strength of interquark potential $H_0$ defined in (2.12) as a function of dual string velocity $v$.  Thus the stable configuration shall satisfy the relation
$$dH_0(v)/dv =0 ~~~\Rightarrow ~~~v^2 = {4B^2\over \sqrt{1+4B^2r_0^2}}. \eqno{(2.13)}$$
Above equation implies that the velocity $v$ of  stable quark-antiquark is an increasing function of the strength of the space non-commutativity.  This property is consistent with the property that the quark-antiquark state in the commutative space is static.   Also, eq.(2.13)  tells us that the velocity $v$ of stable quark-antiquark at a fixed value of non-commutativity is a decreasing function of the radius $r_0$, which also consists with the physical property of a charged particle enters a region with a radius-dependent magnetic field.  Finally, it shall be noticed that as the $B_{\phi z}$ depends on the coordinate $r$ the effect of the non-commutativity on the gauge theory is therefore coordinate dependent.  Thus the interquark potential will depend on  the coordinate $r_0$.

Note that the interquark potential $H(L)$ of arbitrary distance $L$ could be obtained from (2.11) while regularize the expression by integrating the energy only up to $U_{max}$ and subtracting the mass of the W-boson, as that discussed in [4].  The figure 2 shows the interquark potential $H(L)$ as the function of distance $L$.   We see that there presents a minimum distance $L_0$ between the quarks and it becomes Coulomb phase potential in IR.  
\\
\\
\scalebox{1}{\hspace{5cm}\includegraphics{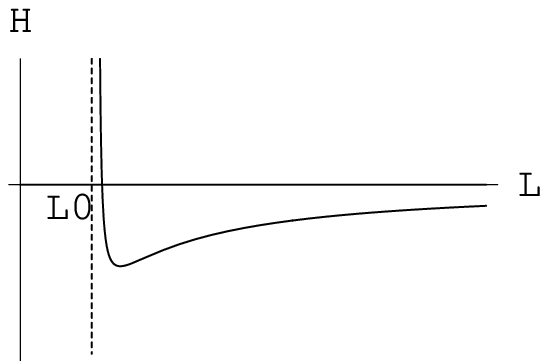}}
\\
\\
{\hspace{3cm} {\it Figure 2.  The interquark potential $H(L)$.  We see that there presents a minimum distance $L_0$ between the quarks and it becomes Coulomb phase potential in IR.}
\\

In this section we have found that the non-constant non-commutativity could produce a strong repulsive force between the quark and anti-quark if they are close enough and there presents a minimum distance between the quarks.  This property does not show in the previous literature in which a constant non-commutativity could only modify the strength of the Coulomb force between the quarks [6,7].    In the next section we will see that the property found in this section could also be shown in the non-commutative dipole theory.

\section{Wilson Loop in Non-commutative Dipole Field Theory}
non-commutative dipole field theory can be regarded as a generalization of ordinary field theory on a non-commutative space which is a non-local theory.  To a dipole field $\Phi_a$ we assign a constant dipole length $\ell_a$ and define the ``dipole product " as [10-12]
$$ \Phi_a(x) \star \Phi_b(x) = \Phi_a (x-{1\over 2}\ell_a)\Phi_b (x-{1\over 2}\ell_b).\eqno{(3.1)}$$
The string theory realization of the non-commutative field theory was found in [13] and geometry is described by
$$ds_{10}^2 = U^2\left(- dt^2+ dx_1^2+ dx_2^2+{ dx_3^2\over 1+\ell^2U^2\sin^2\theta_1\sin^2\theta_2}\right)\hspace{4cm}$$
$$+ {1\over U^2} \left(dU^2+ U^2d\Omega_5^2-U^4\ell^2\sin^4\theta_1\sin^4\theta_2 {a_3d\theta_3+a_4d\theta_4+a_5d\theta_5\over 1+U^2\ell^2\sin^4\theta_1\sin^4\theta_2}\right). \eqno{(3.2)}$$
$$e^{2\Phi}= {1 \over  1+ U^4\ell^2\sin^4\theta_1\sin^4\theta_2},~~~
B_{3\theta_i}= - {a_i~U^2\ell^2\sin^4\theta_1\sin^4\theta_2 \over 1+U^2\ell^2\sin^4\theta_1\sin^4\theta_2 },\hspace{1.7cm}\eqno{(3.3)}$$
in which $a_3 \equiv \cos\theta_4 $, $a_4 \equiv - \sin\theta_3\cos\theta_3\sin\theta_4 $, and $a_5 \equiv \sin^2\theta_3\sin^2\theta_4$, where $\theta_i$ are the angular coordinates parameterizing the sphere $S^5$ transverse to the D3 brane.  Thus there is a nonzero B field with one leg along the brane worldvolume and other transverse to it.

  The supergravity description of the  Wilson loop on the  non-commutative dipole field theory had been evaluated in [14].  Parameterizing the string configuration by $\tau=t$, $U=\sigma$, and $x_3=x_3(\sigma)$ it then found that the quark-antiquark distance becomes
$$L = {2\over U_0}\int_1^\infty dy {\sqrt{1+y^2\ell^2U_0^2}\over y^2~\sqrt{y^4 {1+\ell^2U_0^2\over 1+y^2\ell^2U_0^2}-1}}.\eqno{(3.4)}$$
The energy of the system is 
$$H = {U_0\over \pi}\left[\int_1^\infty dy \sqrt{1+\ell^2U_0^2\over 1+y^2\ell^2U_0^2}\left({y^2\over \sqrt{y^4 {1+\ell^2U_0^2\over 1+y^2\ell^2U_0^2}}-1}-1\right)y^\epsilon-1\right].\eqno{(3.5)}$$
Here we have subtracted the infinity coming from the mass of W-boson which corresponding to the string stretching to $U=\infty$.  The analyses in [14] has found that when the distance between quark and anti-quark is much bigger then their dipole size the interquark energy will show a Coulomb type behavior with a small correction form the non-commutativity.

  In this section we would like to supply an analysis to show that there exist a strong force at short distance.  In the case of  large $U_0$, i.e. at short distance, eq.(3.4) could be approximated as
$$L \approx 2\ell \left(1+{1\over 2U_0^2\ell^2}\right)\left[\int_1^\infty dy {1\over y\sqrt{y^2-1}}- {1\over 2U_0^2\ell^2}\int_1^\infty dy {1\over y^3\sqrt{y^2-1}}\right]= \ell \pi + {\pi\over 4U_0^2\ell}.\eqno{(3.6)}$$
Thus there exist a minimum distance
$$L_0 \equiv \ell \pi,\eqno{(3.7)}$$
between the quarks in the non-commutative dipole theory.  In fact,  performing the numerical evaluation of (3.4) we could obtain the function $L(U_0)$ which is like that plotted in figure 1.  Now, using the relation (3.6) and from (3.5) we can find the interquark potential 
$$H(L) \approx {1\over 2\pi}\sqrt{\pi\over 4(L-L_0)\ell}, ~~~~~as~~ L~\rightarrow~L_0.\eqno{(3.8)}$$
Thus the non-commutativity in the dipole theory may produce a strong repulsive force between the quark and anti-quark if they are close enough.   Eq (3.6) shows that there presents a minimum distance between the quarks and that the distance is proportional to the value of the non-commutativity, exhibiting the nature of the non-commutative theory.

When the distance between quark and antiquark is much bigger than their dipole size the interquark potential calculated in [14]  is given by 
$$H(L) \approx ~-~ {1\over L}\left(1+ {c_0 \ell^2\over L^2}\right),  ~~~~~as~~~L \gg L_0,\eqno{(3.9)}$$
in which $c_0$ is a numerical constant (Note that $L_0\equiv \ell\pi$.).

From (3.8) and (3.9) we see that there has a strong repulsive force between the quark and anti-quark if they are close enough and it becomes Coulomb force in IR, likes that plotted in figure 2.

\section{Conclusion}
In this paper we study the Wilson loop in the some non-commutative gauge theories by using the dual string description in which the corresponding string is on the curved background with B field.   The previous literatures had studied the theory in which a constant B field is turned on along the brane worldvolume and found that the Wilson loop always shows a Coulomb phase.  We have extended the examination to the theory with a non-constant B field, which duals to the gauge theory with non-constant non-commutativity and have found that, while the non-commutativity could modify the Coulomb type potential in IR it may produce a strong repulsive force between the quark and anti-quark if they are close enough.  In particular, we show that there presents a minimum distance between the quarks and that the distance is proportional to the value of the non-commutativity.  We also re-examine the theory in the presence of a nonzero B field with one leg along the brane worldvolume and other transverse to it, which duals to a non-commutative dipole theory.  It also find that the non-commutativity could modify the Coulomb type potential in IR and it may produce a strong repulsive force between the quark and anti-quark if they are close enough.  There also exists a minimum distance between the quarks and that the distance is proportional to the value of the non-commutativity, exhibiting the nature of the non-commutative theory.
\\

Finally, let us make following comments to conclude this paper.

1.  In this paper the expectation value of the Wilson loop is calculated to the lowest order by evaluating the area of the classical string worldsheet.  However, 
since the Melvin background breaks supersymmetry there could have lots of quantum corrections [19-21] which may dramatical change the result. Therefore it is important to study the effects of the worldsheet fluctuations on the expectation value of the Wilson loop. 

2. It is known that, Maldacena method [4] cannot obtain the subleading  corrections which arise when one considers coincident Wilson loops, multiply wound Wilson loops or Wilson loops in a higher dimensional representation. In recent, Drukker and Fiol [22] showed a possible way  to compute a class of these loops using D branes carrying a large fundamental string charge dissolved on their worldvolume pinching off at the boundary of the $AdS$ on the Wilson loop.  It is interesting to use the D-brane approach to evaluate the Wilson loop in non-commutative gauge theory.   Note that the Nambu-Goto action of (2.4) is describing a fundamental string and NSNS B-field effect  is shown in the term $\sim B_{\mu\nu}\epsilon^{ab}\partial_\tau X^{\mu} \partial_\sigma X^{\nu}$.   However, in the D-brane approach the NSNS B-field effect shall be  shown  in the action$\sim \int\sqrt{det(g_{ab}+B_{ab})}$ and, furthermore, there are the Wess-Zumino terms  $\int P[C_4]$ and $\int B\wedge P[C_2]$, in which P[...] denotes the pullback of the RR potential.  

The problems of evaluating the one-loop correction to the classical area of the  worldsheet and using the D-brane approach to evaluate the Wilson loop in non-commutative gauge theory will be investigated in our future study.

~
\\
~
\\
~
\\
~
\\
~
\\
{\bf  \Large References}
\begin{enumerate}
\item J.~M. Maldacena, ``The large {N} limit of superconformal field theories  and supergravity,''  Adv. Theor. Math. Phys.  2  (1998) 231-252  [hep-th/9711200].
\item E.~Witten, ``Anti-de Sitter space and holography,'' Adv.\ Theor.\ Math.\ Phys.\   2 (1998) 253 [hep-th/9802150].
\item S.~S.~Gubser, I.~R.~Klebanov and A.~M.~Polyakov, ``Gauge theory correlators from non-critical string theory,'' Phys.\ Lett.\ B 428 (1998) 105
[hep-th/9802109].
\item J.~M. Maldacena,  ``{W}ilson loops in large {N} field theories,''  Phys.   Rev. Lett.  80 (1998) 4859-4862 [hep-th/9803002]; S.-J. Rey and J.-T. Yee,  ``Macroscopic strings as heavy quarks in large  {N} gauge theory and anti-de {S}itter  supergravity,''   Eur. Phys. J.   C22 (2001) 379--394 [hep-th/9803001]; Y. Kinar, E. Schreiber, and J. Sonnenschein, ``$Q \bar{Q}$ Potential from Strings in Curved Spacetime - Classical Results," Nucl.Phys. B566 (2000) 103-125 [9811192]; J. Gomis and F. Passerini ``Holographic Wilson Loops," JHEP 0608 (2006) 074 [hep-th/0604007].
\item A. Hashimoto and N. Itzhaki,``Non-Commutative Yang-Mills and the AdS/CFT Correspondence," Phys.Lett. B465 (1999) 142 [hep-th/9907166]. 
\item J. M. Maldacena and J. G. Russo,`` Large N Limit of Non-Commutative Gauge Theories," JHEP 9909 (1999) 025 [hep-th/9908134]; U. H. Danielsson, A. Guijosa, M. Kruczenski, and B. Sundborg,``D3-brane Holography," JHEP 0005 (2000) 028 [hep-th/0004187]; S. R. Das and B. Ghosh,``A Note on Supergravity Duals of Noncommutative Yang-Mills Theory," JHEP 0006 (2000) 043 [hep-th/0005007].
\item M. Alishahiha, Y. Oz, and M. M. Sheikh-Jabbari,``Supergravity and Large N Noncommutative Field Theories," JHEP 9911 (1999) 007 [hep-th/9909215].
\item A. Hashimoto and K. Thomas, ``Dualities, Twists, and Gauge Theories with Non-Constant Non-Commutativity," JHEP 0501 (2005) 033 [hep-th/0410123]; A. Hashimoto and K. Thomas, ``Non-commutative gauge theory on D-branes in Melvin Universes," JHEP 0601 (2006) 083 [hep-th/0511197].
\item M. Alishahiha, B. Safarzadeh, and H. Yavartanoo``On Supergravity Solutions of Branes in Melvin Universes," JHEP 0601 (2006) 153 [hep-th/0512036];  Rong-Gen Cai and  N. Ohta, ``Holography and D3-branes in Melvin Universes," Phys.Rev. D73 (2006) 106009 [hep-th/0601044].
\item A. Bergman and O. J. Ganor,``Dipoles, Twists and Noncommutative Gauge Theory," JHEP 0010 (2000) 018 [hep-th/0008030].
\item  K. Dasgupta and M. M. Sheikh-Jabbari, ``Noncommutative Dipole Field Theories," JHEP 0202 (2002) 002 [hep-th/0112064].
\item Wung-Hong Huang,``Exact Wavefunction in a Noncommutative Dipole Field Theory,"  Frontiers In Field Theory, 1-8, Kovras, O. (Editor) 2005, Nova Science Publishers [hep-th/0208199].
\item A. Bergman, K. Dasgupta, O. J. Ganor, J. L. Karczmarek, and G. Rajesh,``Nonlocal Field Theories and their Gravity Duals," Phys.Rev. D65 (2002) 066005
[hep-th/0103090].
\item M. Alishahiha and H. Yavartanoo,``Supergravity Description of the Large N Noncommutative Dipole Field Theories," JHEP 0204 (2002) 031 [hep-th/0202131].
\item J.~M.~Maldacena and C.~N\'u\~nez, ``Towards the large N limit of pure N = 1 super Yang Mills,'' Phys.\ Rev.\ Lett.\   86 (2001) 588 [hep-th/0008001].
\item T. Mateos, J. M. Pons, P. Talavera,  ``Supergravity Dual of Noncommutative N=1 SYM,'' Nucl.Phys. B651 (2003) 291-312 [hep-th/0209150]. 
\item Wung-Hong Huang, ``AdS/CFT Approach to Melvin Field Deformed Wilson Loop," [hep-th/0612018].
\item F.~Dowker, J.~P.~Gauntlett, D.~A.~Kastor and J.~Traschen, ``The decay of magnetic fields in Kaluza-Klein theory,'' Phys.\ Rev.\ D52 (1995) 6929 [hep-th/9507143]; M.~S.~Costa and M.~Gutperle, ``The Kaluza-Klein Melvin solution in M-theory,'' JHEP 0103 (2001) 027 [hep-th/0012072]; Wung-Hong Huang, ``Semiclassical Strings in Electric and Magnetic Fields Deformed $AdS_5 \times S^5$ Spacetimes,'' Phys.Rev. D73 (2006) 026007 [hep-th/0512117].
\item  J. Greensite and P. Olesen, ``Worldsheet Fluctuations and the Heavy Quark Potential in the AdS/CFT Approach," JHEP 9904 (1999) 001 [hep-th/9901057];  S. Forste, D. Ghoshal, and S. Theisen, ``Stringy Corrections to the Wilson Loop in N=4 Super Yang-Mills Theory," JHEP 9908 (1999) 013 [hep-th/9903042].
\item  Y. Kinar, E. Schreiber, J. Sonnenschein, and N. Weiss.``Quantum fluctuations of Wilson loops from string models," Nucl.Phys. B583 (2000) 76 [hep-th/9911123].
\item  N. Drukker, D. J. Gross, and A. Tseytlin,`` Green-Schwarz String in $AdS_5 \times S^5$: Semiclassical Partition Function," JHEP 0004 (2000) 021 [hep-th/0001204];  K. Zarembo, ``Supersymmetric Wilson loops," Nucl.Phys. B643 (2002) 157 [hep-th/0205160].
\item N.~Drukker and B.~Fiol, ``All-genus calculation of Wilson loops using
  D-branes,''  JHEP  02 (2005) 010 [hep-th/0501109]; S.~A. Hartnoll and S.~Prem~Kumar,  ``Multiply wound Polyakov loops at strong coupling,''  Phys. Rev.  D74 (2006) 026001[hep-th/0603190]; S. Yamaguchi, ``Wilson loops of anti-symmetric representation and  D5-branes,''  JHEP 05 (2006) 037 [hep-th/0603208].
\end{enumerate}
\end{document}